\documentclass[aps,prl,twocolumn,showpacs,10pt,amsmath,footinbib]{revtex4-1}
\usepackage{graphicx}
\usepackage[dvipsnames]{xcolor}
\usepackage{dcolumn}
\usepackage{bm}
\usepackage{hyperref}
\definecolor{red}{RGB}{240,34,31}
\definecolor{green}{RGB}{50,200,50}
\definecolor{darkergreen}{RGB}{20,100,20}
\definecolor{blue}{RGB}{20,20,240}

\begin{document}
\title{A superfluid boundary layer}

\author{G. W. Stagg}
\email{george.stagg@newcastle.ac.uk}
\author{N. G. Parker}%
\author{C. F. Barenghi}
\affiliation{Joint Quantum Centre (JQC) Durham-Newcastle, School of Mathematics and Statistics, Newcastle University, Newcastle upon Tyne, NE1 7RU, UK
}

\date{\today}%

\begin{abstract}
We model the superfluid flow of liquid helium over 
the {rough} surface of a wire (used to experimentally generate turbulence)
profiled by atomic force microscopy.    
Numerical simulations of the Gross-Pitaevskii equation reveal that 
the sharpest features in the surface induce vortex nucleation 
both intrinsically 
(due to the raised local fluid velocity) and extrinsically ({providing}
pinning sites to vortex lines aligned with the flow). 
Vortex interactions and reconnections contribute to form a  
dense turbulent layer of vortices with a non-classical average velocity 
profile which continually sheds small vortex rings into the bulk.   
We characterise this layer for various imposed flows. 
As boundary layers conventionally arise from viscous 
forces, this result 
%is surprising and 
opens up new insight into the 
nature of superflows.
% at \red{rough} boundaries. %600 chars
\end{abstract}

%some superfluid, 3He and 4He and vortex pacs
%\pacs{67.30.H-, 67.30.he, 67.25.D-, 47.37.+q, 03.75.Lm}

\maketitle

%%%%%%%%%%%%%%%%%%%%%%%%%%%%%%%%%%%%%%%%%

At sufficiently low temperatures, liquid helium has two striking
properties.  Firstly, it flows without viscosity.  Secondly,
its vorticity is constrained to
thin mini-tornadoes, characterised by fixed 
circulation $\kappa$ (the ratio of Planck's constant to
the mass of the relevant boson - one atom in $^4$He and one Cooper pair 
in $^3$He-B) and microscopic core radius $\xi$ 
($0.1~\rm nm$ in $^4$He and $10~\rm nm$ in $^3$He-B).  
In contrast, the eddies in everyday viscous fluids can have arbitrary shape, 
size and circulation.  

Of ongoing experimental and theoretical study is the nature of 
turbulence in superfluids 
\cite{Barenghi-PNAS,Bradley2011-NatPhys,Zmeev2015a,Boue2013}, a state 
consisting of an irregular tangle of quantised vortex lines.  
Despite fundamental differences between superfluids and classical fluids, 
the observations of Kolmogorov energy spectra (famed from classical 
isotropic turbulence) in superfluid turbulence \cite{Barenghi-PNAS} 
are suggestive of a deep connection between them.  
Superfluid turbulence is nowadays most commonly formed by moving obstacles, 
including grids \cite{Davis2000}, 
wires \cite{Guenault1986,Bradley2005,Bradley2011,Fisher2001}, 
forks \cite{Blaauwgeers2007,Bradley2012}, 
propellers \cite{Tabeling1998,Salort2010}, spheres \cite{Schoepe1995} 
and other objects \cite{VinenSkrbek2008}.
Despite progress in visualizing the flow of superfluid helium in the
bulk \cite{Zmeev2015b,Duda2015}, including individual vortex reconnections
\cite{Lathrop}, {the study of flow profiles 
\cite{Xu2007,Guo2014} is still in its infancy} and
there is {no direct} experimental evidence about what happens 
at boundaries.
Here, 
vortices are believed to be generated by two mechanisms.  Firstly, 
 {vortices can nucleate} at the boundary of the 
vessel or object \cite{Pomeau1992}.
When the relative flow speed is sufficiently low, the flow is laminar 
(potential) and dissipationless.  
Near curved boundaries, however,  {intrinsic} vortex nucleation occurs if
the local flow velocity exceeds a critical value. 
Secondly, the vortices can be procreated (extrinsically generated)
{by the  `vortex-mill' mechanism\cite{Schwarz-mill} from}
so-called `remanent vortices' 
which are present in the system since cooling the helium through the 
superfluid transition.  Remanent vortices can be avoided using judicious, 
slow experimental protocols \cite{Yano-2007}. 
 
The nano-scale vortex core in superfluid helium is comparable in size 
to the typical roughness of the boundaries of the vessel or stirring object. 
Unfortunately the lack of direct experimental information about vortex 
nucleation at the boundaries and the subsequent vortex-boundary interactions,
limit the interpretation of experiments. Theoretical
progress is challenging and to date has focussed on smooth and 
idealised surfaces.  In principle, the superfluid boundary conditions
are straightforward: the superfluid velocity
component which is perpendicular to the boundary must vanish
at the boundary, whereas the tangential component (in the absence of
viscous stresses) can slip.  For the latter reason, 
{in superfluids} we do not expect 
boundary layers typical of viscous flows.   
Implementing these superfluid boundary conditions, 
it was found\cite{Schwarz-1981-pinning, Tsubota-1994-pinning}
that one or more vortices sliding along a smooth surface
can become deflected or trapped by small
hemispherical bumps.  Such bumps can also serve as nucleation sites 
for vortices; the local superfluid velocity is raised at the pole 
of the bump and more readily breaks the critical velocity for 
vortex nucleation\cite{winiecki}.  
Indeed, our recent simulations\cite{Stagg-ellipse} have shown that, if
the bump is elliptically shaped and elongated perpendicular to the 
imposed flow, the superfluid velocity $v$ at the pole
is enhanced,  {reducing the critical Mach number for
vortex nucleation from $v/c \sim 1$ to smaller values}
$v/c \sim \epsilon^{-1} \ll 1$ (where $\epsilon \gg 1$ is the ellipticity of 
the bump), and increasing the {intrinsic} vortex nucleation rate 
(for a given super-critical imposed flow). 
We expect therefore that microscopically-small surface roughness 
may promote the nucleation of vortices at a surface.
For pre-existing vortex lines in the vicinity of the surface, there is also 
indirect experimental evidence of a `vortex mill' mechanism 
which continuously feeds vorticity into the flow by stretching 
 {any pre-existing} vortex lines. This mechanism only works if 
the spooling vortex, 
held by pinning sites at the surface, is aligned in the streamwise 
direction \cite{Schwarz-mill}.   In summary, boundary roughness
potentially affects both intrinsic and extrinsic 
mechanism to create new vortices.

\begin{figure}
\includegraphics[width=0.86\linewidth]{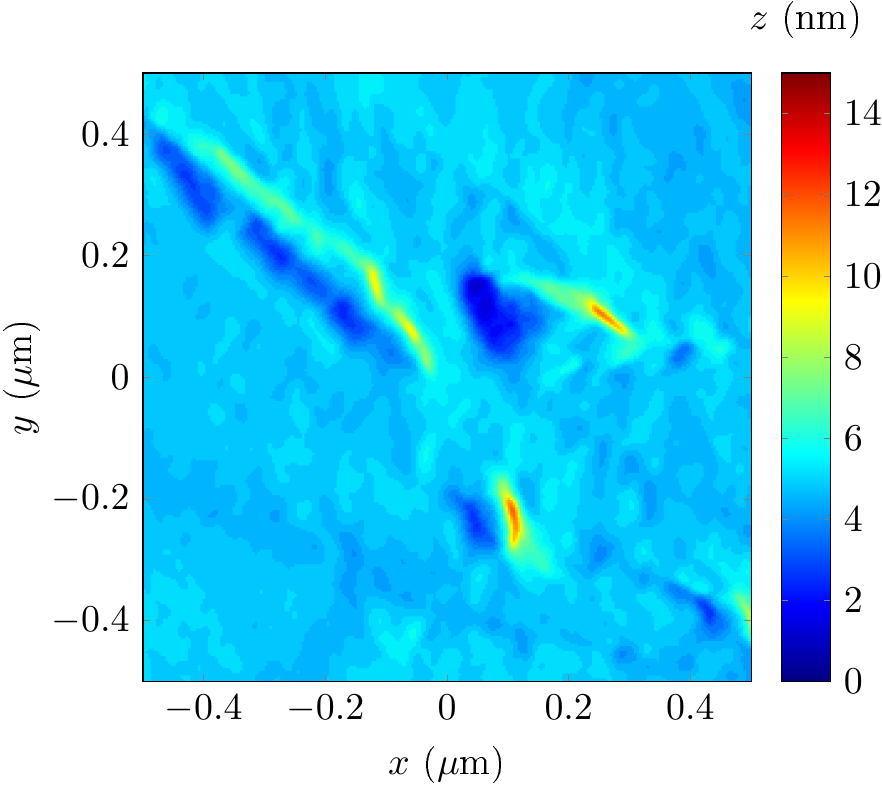}
  \caption{\label{fig:afmsmooth}AFM image of a section of the NbTi wire rough surface, smoothed by a Gaussian blur (standard deviation 6 nm) so as to remove discontinuities in the surface profile.}
 \end{figure}

To shed light on the problem,
% microscopic behaviour of superflow near a rough boundary, 
we work with the 3D profile of a {rough} surface 
[Fig.~\ref{fig:afmsmooth}].  This corresponds to a  ($1 \mu$m)$^2$ region 
of the surface of a thin NbTi wire used to generate quantum turbulence 
at Lancaster University, as profiled via atomic force microscopy 
(AFM) \cite{Lawson}.  The surface is rough, with a height up to around 
10 nm, and features sharp grooves and steep ridges, likely to have 
arisen during the etching phase of the wire preparation.    
We assume that such a `mountain'  landscape is typical of the wires 
and similar objects used in experiments.

We model the flow of
superfluid helium over this surface  
through the time-dependent Gross-Pitaevskii 
equation (GPE) for a weakly-interacting Bose superfluid \cite{SM}.  
The GPE describes a fluid, of density $n({\bf r},t)$ and velocity 
${\bf v}({\bf r},t)$, which follows a classical 
continuity equation and a modified Euler equation (the modification being 
the presence of a quantum pressure term, arising from zero-point motion 
of the particles and responsible for vortex nucleation and reconnections).  
While the GPE provides only a qualitative model of the strongly-interacting 
superfluid helium (for example, the GPE's excitation spectrum lacks 
helium's roton minimum), it nevertheless contains the key microscopic 
physical ingredients of our problem: finite-size vortex core, 
vortex interactions and vortex reconnections.  
The more traditional vortex filament model \cite{Schwarz1988},
used to model the motion of vortex lines in the presence of
smooth spherical \cite{Hanninen-sphere,Kivotides-sphere},
hemispherical \cite{Schwarz-1981-pinning,Tsubota-1994-pinning} 
and cylindrical boundaries \cite{Hanninen-PNAS,Goto2008}, 
is less appropriate for a number of reasons: it assumes that the vortex core is infinitesimal compared to 
any other length scale (which is not the case if 
vortex core and wall roughness are comparable); it does not 
contain vortex nucleation and kinetic energy losses due
to sound emission; and it is difficult to 
generalise from smooth, geometrically simple (cylindrical or spherical)
boundaries to rough boundaries.

The bulk fluid has uniform average density $n_0$, with the surface imposed 
as an impenetrable region.  The characteristic scales of length and speed are 
healing length $\xi$ (the vortex core size) and speed of sound $c$, respectively.
A characteristic time scale follows as $\tau=\xi/c$.
We simulate the superfluid flowing at an imposed speed $v$ over the entire 
AFM surface, in a 3D domain, periodic in $x$ and $y$.  
The surface area ($1 \mu$m$)^2$ is mapped onto the largest practical 
healing length area of $(400 \xi)^2$ \cite{SM}.

\begin{figure}
\includegraphics[width=0.93\linewidth]{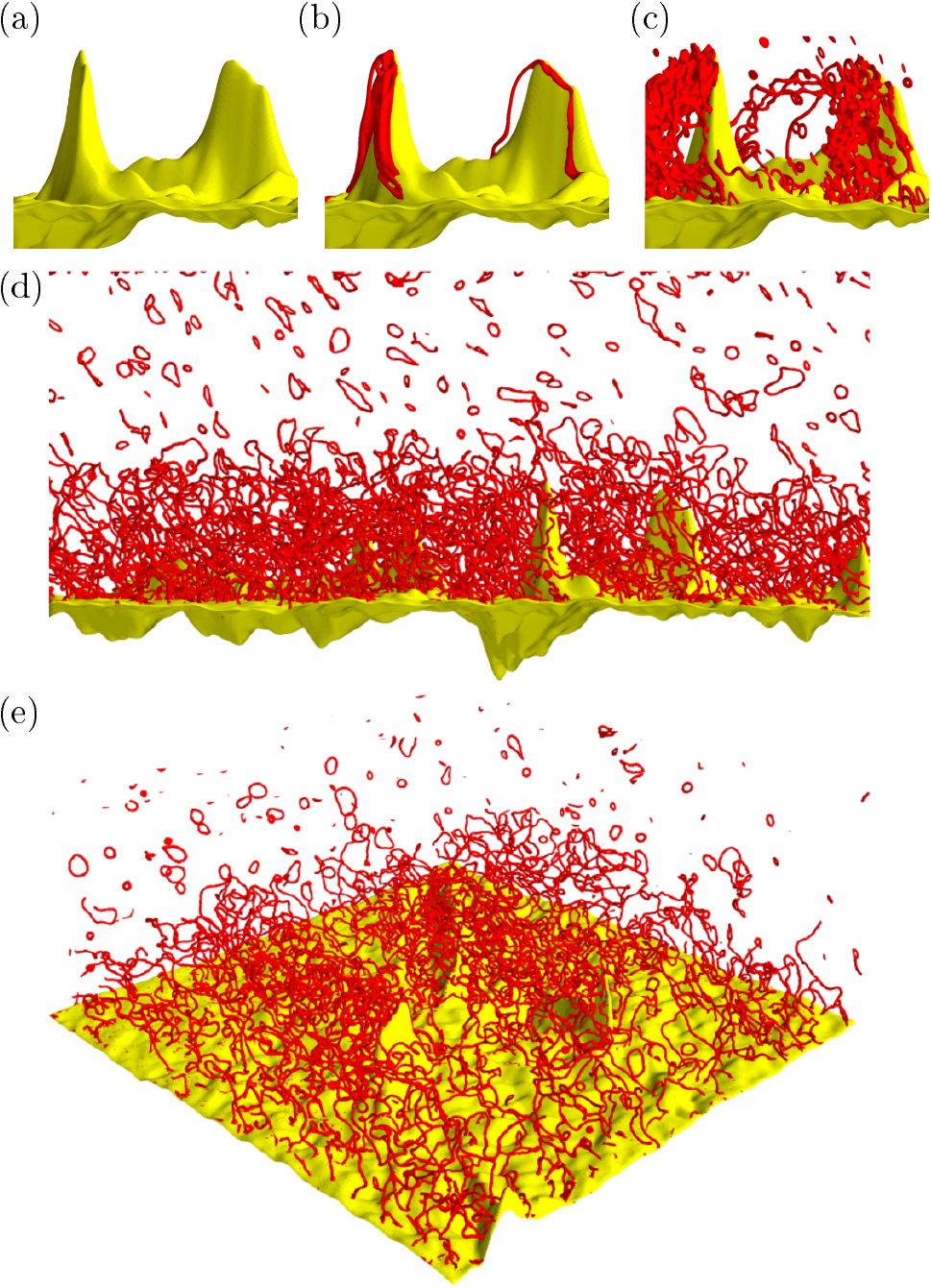}
  \caption{\label{fig:zoomiso}  Vortex nucleation and formation of the 
turbulent boundary layer for imposed flow $v=0.6\,c$.  
(a-c)  Isosurface density plots ($0.25n_0$), showing the surface (yellow) 
and vortices (red) in the vicinity of the two tallest mountains 
(view taken along $y$ for 
$15 \xi \leq x \leq 125 \xi$) at times 
$t=20,30,100\,\tau$.  In (c) note three vortex lines which are aligned
along the imposed flow and develop unstable Kelvin waves which will
reconnect  and create new vortex loops.
(d-e) Isosurfaces of the entire surface in the saturated turbulent 
regime at late times ($t=1220 \tau$).  Note the turbulent 
layer up to approximately the height of the tallest mountains 
and the region of small vortex rings above it.}
\end{figure}

In the vicinity of the surface the local fluid speed 
is enhanced by the surface's roughness, 
with the maximum values occurring near the tallest mountains.   
Up to a critical speed, $v_{\rm c}$, the flow remains vortex-free.  
For increased imposed flow velocity, $v_{\rm c}$ is 
first exceeded at the highest mountain,  leading to vortex 
nucleation [Fig. \ref{fig:zoomiso} (a-c)], and then at other 
high mountains on the surface.  The critical velocity for vortex 
nucleation across this surface occurs for an imposed 
flow $v_{\rm c}\approx 0.2 c$; this is considerably smaller than, say a 
hemispherical bump for which $v_{\rm c} \approx 0.5 c$ \cite{winiecki},  
indicating the significant role of the surface roughness in enhancing 
the breakdown of laminar flow.  

We focus on the imposed flow speeds $v=0.3\,c, 0.6\,c$ and
$0.9\,c$, each well exceeding $v_{\rm c}$.
Nucleated vortices either peel off the boundary or, more frequently, slide
down the slopes of the mountains in the form of partially attached 
vortex loops (carried by the imposed flow). Nucleated vortex loops 
are of the same circulation and form clusters (manifesting as 
partially attached vortex bundles) on the leeward side of the mountains,
{see Fig.~\ref{fig:zoomiso}(b)}.
The velocity field of vortex bundles and the nucleation 
of small vortex loops throughout the surface 
cause vortex stretching and reconnections, 
distorting the bundles of vortices and small rings into a 
complex tangle downstream of the mountains.
The tangle 
is continuously fed by further vortices which are generated.
The formation of the tangle is shown 
in Fig.~\ref{fig:zoomiso}(c), and the fully developed 
turbulent layer near the surface is seen
in Fig.~\ref{fig:zoomiso}(d-e).

As the number of vortices 
increases, the turbulent region remains strongly 
localised near the surface, up to approximately the 
height of the tallest mountain, forming a distinct layer [Fig.~\ref{fig:zoomiso}(d)].  
Vortex reconnections 
cause a continuous ejection of vortex rings which spread into the bulk 
[Fig.~\ref{fig:zoomiso}(d, e)]. These small rings,
predicted by\cite{Kursa2011,Kerr2011}, play an important role in turbulent 
cascades\cite{Svistunov1995}.

\begin{figure}[t]
  \centering
\includegraphics[width=0.95\linewidth]{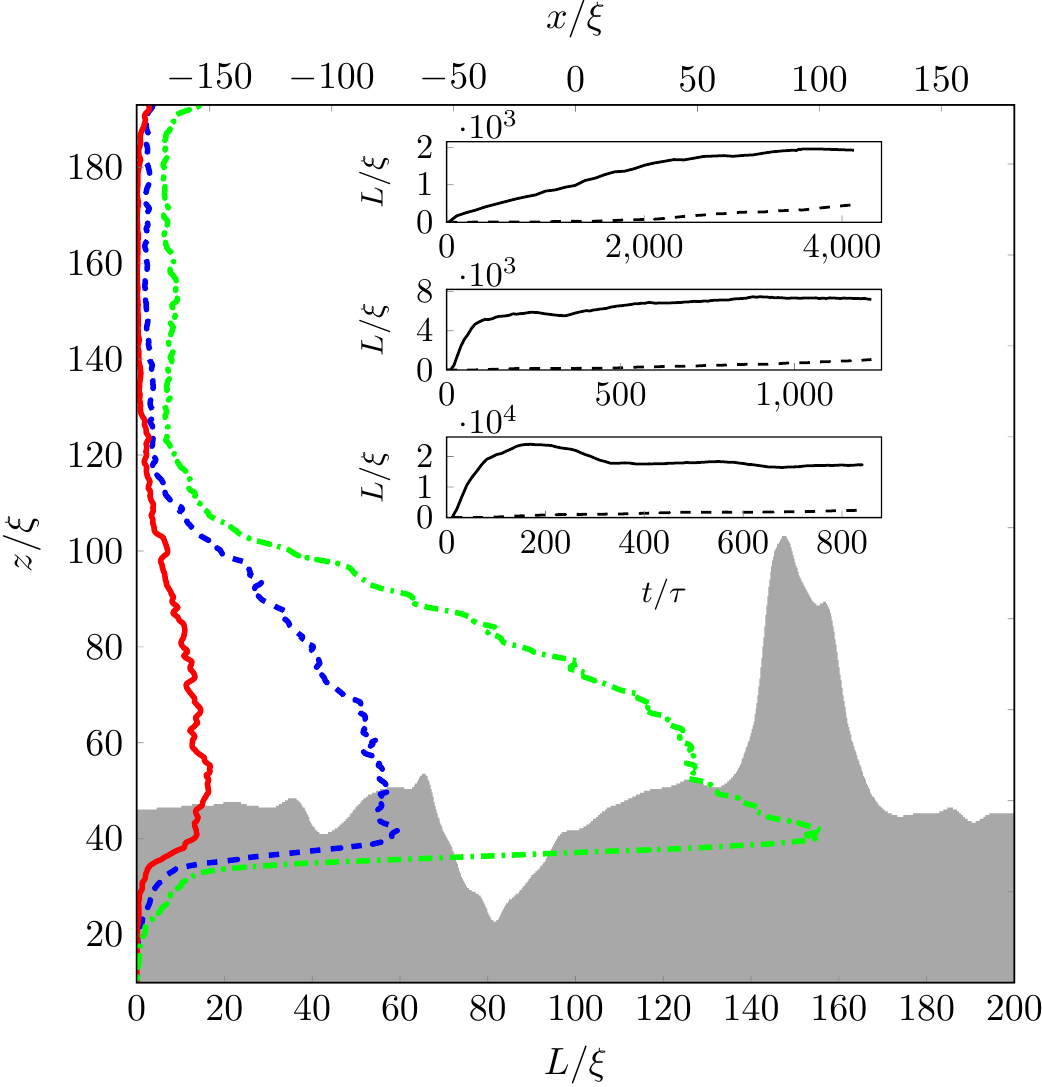}
  \caption{\label{fig:line-length}  {Average vortex line length, $L$ (bottom scale), 
  as a function of  height, $z$ (left scale), for $v=0.3\,c$ (solid red line), $v=0.6\,c$ 
  (dashed blue line), and $v=0.9\,c$ (dot-dashed green line) 
in the saturated regime. 
A 2D slice ($y=0.1\,\mu$m) of the 3D surface along $x$ (top scale) is shown in grey to 
visualize the height of the highest mountains. Inset: Vortex line length below 
($L_0$, {solid line}) and above ($L_1$, {dashed line}) 
the height of $z=100\xi$ (approximately the height of the 
highest mountain) for imposed flow speeds
$v=0.3\,c$ (top), $0.6\,c$ (middle) and $0.9\,c$ (bottom).}
}
\end{figure}

\begin{figure}
\begin{center}
\includegraphics[width=0.8\linewidth]{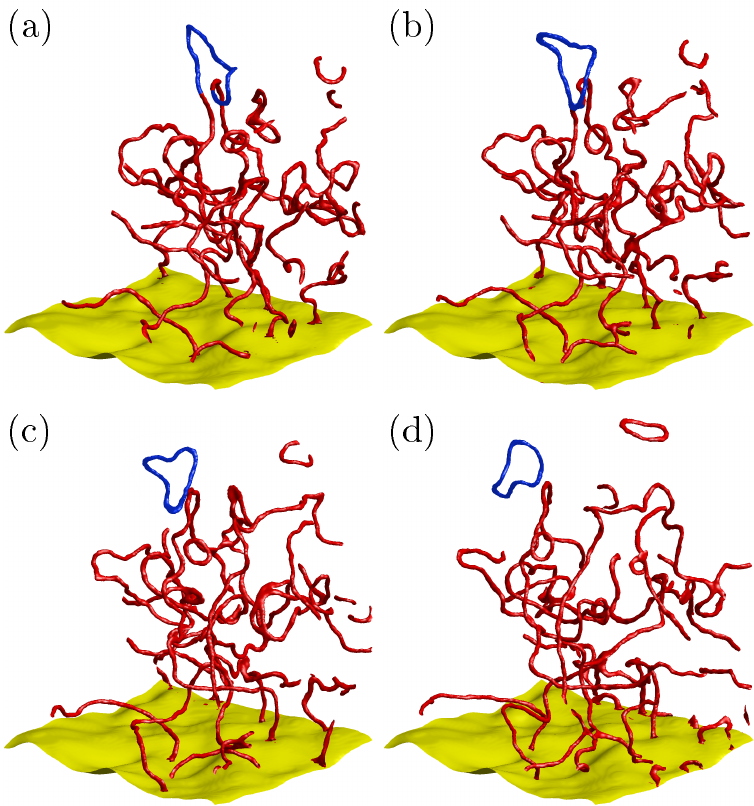}
\end{center}
\caption{Extrinsic nucleation of a vortex ring (highlighted in blue) 
from the boundary layer, which escapes into the bulk.  
As for Fig.~\ref{fig:zoomiso} but zoomed up on a ($76\xi)^2$ region of 
the surface at times $t=730, 640, 750$ and $770~\tau$.}
\label{fig:escapering}
\end{figure}

The turbulent layer and ejected vortex rings are not isotropic: 
on average, vortex lines tend to be flattened, parallel to the surface;
the ejected rings also tend to lie more in the $xy$ plane (and 
travel vertically away from the layer).

We monitor the vortex line length below the tallest mountain 
($z \approx 100 \xi$), $L_0$, and above it, $L_1$ 
[Fig.~\ref{fig:line-length} (inset)].
For $v=0.6c$,  $L_0$ increases with time and saturates.  
Meanwhile, $L_1$ rises slowly, as small rings are continually 
shed by the turbulent layer into the bulk. Repeating for  slower ($v=0.3c$) 
and  faster 
($v=0.9c$) imposed flows reveals the same qualitative behaviour, 
but where the layer forms at a slower and faster rate, respectively.
The resulting vortex line length distribution is shown in Fig.~\ref{fig:line-length}.
The vortices are predominately located near the surface of the wire, with a faster 
imposed flow leading to denser turbulent layers.

At early times, vortex lines which become aligned along the flow direction 
may twist and generate further vortices. 
Surface roughness favours this effect by providing pinning sites
for streamwise-aligned vortices  {which develop Kelvin waves and reconnect,
spooling new vorticity.
An example 
of this vortex-mill mechanism \cite{Schwarz-mill} 
can be seen in Fig. \ref{fig:zoomiso} (c). 
This confirms that %, for the AFM surface and an imposed flow of $u=0.6\,c$, 
the vortex tangle which develops can be interpreted as generated either 
intrinsically, or extrinsically by the vortex-mill
mechanism: 
in both cases vortices nucleate at the tallest mountains before filling 
the layer below.}

At later times (when the turbulent layer of vortices has
saturated) and/or for higher imposed flow velocities, the critical velocity is 
exceeded across greater areas of the surface. However,  
the highest mountains continue to dominate vortex generation; here the fluid 
velocity is always the highest and vortex shedding occurs at the fastest rate. 
To maintain equilibrium, vortex line-length is continuously ejected from the 
top of the turbulent layer by vortex twisting and
reconnections which create small vortex rings that detach and travel 
upwards in the positive $z$ direction. An example is seen 
in Fig.~\ref{fig:escapering}, and highlights the role of
reconnections (hence of the quantum pressure)
in creating new vortices.

\begin{figure}[b]
  \centering
\includegraphics[width=0.95\linewidth]{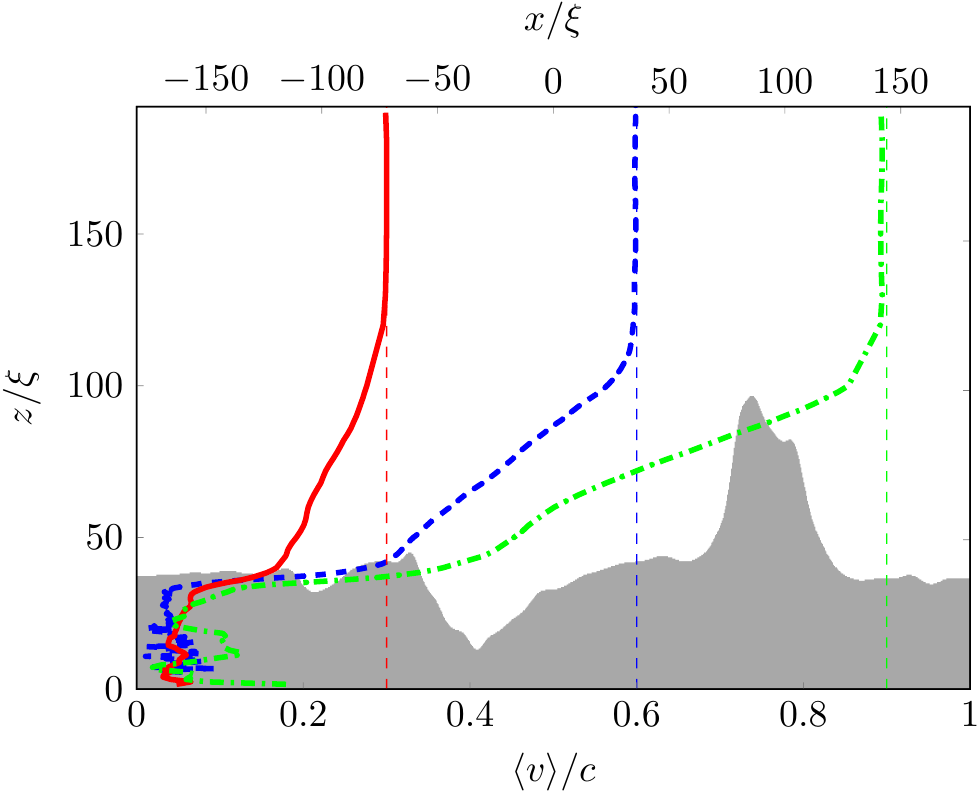}
  \caption{\label{fig:profile}
 {Average superfluid velocity, $\langle v \rangle$ (bottom scale), 
as a function of  height, $z$ (left scale), for
$v=0.3\,c$ (solid red line), $v=0.6\,c$ (solid blue line), and $v=0.9\,c$ (dot-dashed green line) 
in the saturated regime. The grey surface silhouette is as in Fig. \ref{fig:line-length}.
}
}
\end{figure}

To characterise the turbulent layer in a quantitative way,
we determine the average turbulent velocity $\langle v \rangle $ \cite{vz} as a 
function of height $z$ for the three imposed flow speeds [Fig.~\ref{fig:profile}].
In all cases the turbulent layer consists of three regions.
In the top region $100\xi \lesssim z \lesssim 200\xi$, $\langle v \rangle$  
is equal to the velocity of the applied flow,
showing that, above the height of the tallest mountain,
the flow is unaffected by the rough surface underneath.
In the middle region $40 \lesssim z \lesssim 100\xi$, the presence of vortices 
near the surface creates a velocity field that counteracts the 
imposed flow: the closer to the surface one is, 
the slower $\langle v \rangle$ is.
In the bottom region $0 \lesssim z \lesssim 40\xi$, most of the computational 
volume is below the average surface, and only the fluid in the 
valleys contributes to $\langle v \rangle$, which
rapidly drops to near zero.

{The difference between the energy which is fed into the turbulent layer
by the incoming (uniform) flow profile and the energy removed by the 
(approximately linear) profile
is the energy dissipated into sound waves \cite{Parker2005,Leadbeater2001}. 
In classical turbulence, 
the energy dissipation can be related
to the kinematic viscosity $\nu$ of the fluid. 
In our problem, we estimate \cite{SM}
that the emergent $\nu$ 
is $\nu/\kappa \approx 2.4, 1.5$ and $1.1$ at the
three imposed flow speeds, 
larger than 
$\nu/\kappa \approx 0.1$ reported in He$^4$ experiments\cite{Golov2008}.
However, in our problem the vortex lines are
much closer to each other, relative to the vortex core size: the ratio of the
average vortex distance $\delta \approx {\cal L}^{-1/2}$ and vortex 
core radius $a_0$ at the three imposed speeds is
$\delta/a_0 \approx 13.8, 7.4$ and $5.1$, whereas
$\delta/a_0 \approx 2 \times 10^6$ is typical of He$^4$ experiments \cite{SM}. 
Stronger accelerations and more frequent reconnections 
justify the larger dissipation in our problem.
}

{
The analogy with classical fluids was recently pursued by the
introduction of the superfluid Reynolds 
number \cite{Reeves2015,Schoepe2015}
$Re_{\rm s}=(v-v_{\rm c})D/\kappa$ (where $D$ is the length scale of the problem),
Quasi-classical flows (such as Karman
vortex street configurations for the flow past an obstacle \cite{Sasaki2010,
Stagg-ellipse,Reeves2015,Kwon2016}) appear only if $Re_{\rm s}$ is sufficiently large.
2D simulations \cite{Reeves2015} suggest that turbulence onsets 
for $Re_{\rm s} > Re_{\rm c}=0.7$ (the Karman vortex streets becoming irregular).
In our case $D=60 \xi$ and $\kappa=2 \pi c \xi$; for the three applied flow
speeds we find $Re_{\rm s} \approx 1.0, 3.8, 6.7$, all larger than the cited
$Re_{\rm c}$, which is to be expected since we have developed turbulence.
}

In classical fluid dynamics boundary layers 
arise from viscous forces which are absent in low temperature superfluid 
helium.
A classical fluid boundary layer is either laminar or turbulent, and
it is natural to ask whether there is any 
transition from laminar to turbulent boundary layer for our problem.
In classical laminar flow, sheets of fluid slide past
each other, smoothly exchanging momentum and energy only via
molecular collisions at the microscopic scale; in the turbulent case,
eddies induce mixing across sheets which are macroscopically separated from 
each other.  The superfluid analog of laminar flow is 
potential (vortex-free) flow.
% or
%, one can argue, 
%a polarized configuration of vortex
%lines (so that the coarse-grained superfluid
%velocity could be interpreted as a laminar vortex flow). 
Our simulations show
either vortex-free flow or turbulent flows past the rough surface,
so they describe a transition from laminar flow to developed turbulence.
The bottom region of fluid ($0 \lesssim z \lesssim 40\xi$) is a poor 
analog to a classical laminar viscous sub-layer because it contains 
irregular vortex lines which terminate at the boundary.

In conclusion,
our findings illustrate a deep analogy 
between classical and quantum fluids in the presence of boundaries, besides
the analogies already noticed\cite{Barenghi-PNAS}
in homogeneous isotropic turbulence.  
Our results also suggest that the walls which
confine the flow of superfluid helium and the surfaces of moving
objects used to generate turbulence (wires, grids, propellers, spheres)  
may be covered by a thin layer of tangled vortex lines.
The experimental implications of such `superfluid boundary layer' 
on macroscopic observables need to be investigated,
particularly
in $^3$He-B, where, due to relative
large healing length, it is possible to control 
surface roughness.

We thank R. P. Haley and C. Lawson (Lancaster University) for providing the
AFM image of their `floppy wire'. 
C.F.B. acknowledges funding from EPSRC
(grant EP/I019413/1). 
This work used the facilities of N8 HPC, provided and funded by 
the N8 consortium and EPSRC 
(grant EP/K000225/1).

Data supporting this work is openly available under an Open Data Commons Open Database License \cite{data}.

\section*{Supplementary Material}

\subsection*{Gross-Pitaevskii model of the superfluid}
The superfluid is modelled as a weakly-interacting Bose gas.  
It is parameterised by a complex wavefunction 
$\Psi({\bf r},t)=\sqrt{n({\bf r},t)}\exp[i S({\bf r},t)]$, where $n$ and $S$ 
represent the distribution of the particle number density and phase, respectively.
The wavefunction obeys the time-dependent Gross-Pitaevskii 
equation \cite{Pethick,RobertsBerloff-GPE},
\begin{equation}
i \hbar \frac{\partial\Psi}{\partial t} 
= \left(-\frac{\hbar^2}{2m}\nabla^2 + V({\bf r},t) + g|\Psi|^2 \right) \Psi.
\label{eq:gpe}
\end{equation}
where $m$ denotes the particle mass, the coefficient $g$ accounts for local particle interactions, and $V({\bf r},t)$ describes a potential landscape acting on the fluid.  
The GPE is equivalent to a hydrodynamic model with fluid 
density $n({\bf r},t)=|\Psi({\bf r},t)|^2$ and velocity 
${\bf v}({\bf r},t)=(\hbar/m)\nabla S$, and embodies a classical 
continuity equation and a modified Euler equation (the modification being 
the presence of a quantum pressure term, arising from zero-point motion 
of the particles and responsible for vortex nucleation and reconnections).  

Within the GPE model and assuming a homogeneous particle density $n_0$, the vortex core size is characterised by the healing length $\xi=\hbar/\sqrt{m n_0 g}$.  The natural speed, energy and time scales are provided by the speed of sound $c=\sqrt{n_0 g/m}$, the chemical potential $\mu=n_0 g$ and the unit $\tau=\xi/c$, respectively.  

\subsection{Set-up of the simulations}

Our results are based on simulations of the GPE over the entire AFM surface, resolved down to a sub-core scale of $\Delta=0.4\xi$.  In $^4$He the vortex core size is $a_0 \approx 10^{-10}~\rm m$ \cite{Rayfield1964}, such that the $(1\mu{\rm m})^2$ AFM image has true core dimensions $(10^4 \times 10^4 \times 100) a_0^3$.  It is not computationally feasible to model the corresponding range of scales directly within the GPE; as such we map the AFM image onto the largest practical healing length volume of $(400 \times 400 \times 100) \xi^3$.  This is simulated in a box of size $(400 \times 400 \times 200) \xi^3$ (the numerical domain being twice as high as the highest mountain in the third dimension), on a $1000\times 1000\times 500$ spatial grid, which is periodic in $x$ and $y$.  Time evolution is performed with 4th order Runge-Kutta scheme
with time step $\Delta t=0.01 \tau $, and performed across 256 (2.6 GHz) cores of a computer cluster.  The rough surface is incorporated into the GPE by setting a potential barrier $V=5\mu$ below the surface, heavily prohibiting density there; above the surface the density recovers to the bulk value $n_0$ (where $V=0$) over a lengthscale of the order of $\xi$.

We first obtain the stationary solution of the GPE in the presence of
the rough boundary, with bulk density $n_0$, by solving the GPE in imaginary time \cite{Primer}.  The GPE is then transformed into a frame moving at speed $u$ in the $x$-direction (corresponding to the imposed flow) by the addition of a Galilean boost term $ i \hbar u \partial \Psi/\partial x$ to the right-hand side of the GPE.  The flow speed {is}
ramped up smoothly from zero to the required final value. 

%\section{Superfluid Reynolds number}
%
%\red{
%The superfluid Reynolds number\cite{Reeves2015} is
%}
%
%\begin{displaymath}
%Re_s=\frac{(v-v_c) D}{\kappa},
%\end{displaymath}
%
%\noindent
%\red{
%where $D$ is the relevant length scale of the problem, $\kappa$ the
%quantum of circulation,  $v_c$ the critical velocity of vortex
%nucleation and $v$ the speed of the flow.  In our work, $D \approx 60 \xi$,  
%$\kappa=2 \pi$, $v_c=0.2 c$, and $v=0.3c, 0.6c, 0.9c$. 
%}

\subsection{Emergent kinematic viscosity}

{
We estimate the effective kinematic viscosity in the following way. 
Consider the volume $V=D^2 h$ where $D=200 \xi$ is the extension in the
$x$ and $y$ direction and $h \approx 60 \xi$ is approximately the height of the 
turbulent layer in the $z$ direction.  The energy that the uniform flow 
of speed $v$ (along the $x$ direction) brings into this volume
in time $\Delta t$ is $E_{\rm in} = \rho D h v^3 \Delta t/2$. In the turbulent
layer, the velocity is not uniform, but is approximately $v z/h$. The energy
which this shear flow takes out of the said volume is 
$E_{\rm out}=\rho D v^3 \Delta t/8$. In time $\Delta t$, the energy difference
$E_{\rm in}-E_{\rm out}$ is deposited into the turbulence. Since we are dealing with
a statistical steady state, this energy is also dissipated into sound waves
when vortices accelerate around each other \cite{Parker2005}
or reconnect \cite{Leadbeater2001} (including reconnections with images
across the rough surface). For
the sake of simplicity, we group the
small vortex rings which are emitted by the turbulent layer with the
sound waves.
}

{
In time $\Delta t$, the energy (per unit mass)  
dissipated by the turbulence is thus $\Delta E'=(3/8 h) v^3 \Delta t$.
Following Walmsley and Golov's \cite{Golov2008} analysis of their
experiments in low-temperature He$^4$, we use the classical
result that relates the rate of energy dissipation (per unit mass), $dE'/dt$, 
to the kinematic viscosity $\nu$ and the vorticity $\omega$ of the fluid,
}
\begin{equation}
\frac{dE'}{dt}=- \nu \omega^2.
\end{equation}

\noindent
{
We also make the identification \cite{Golov2008}
$\omega \approx \kappa {\cal L}$ where ${\cal L}$ is the vortex
line density, defined as the vortex length $L$ per unit volume $V$
(for the sake of simplicity, we neglect the volume of the
`mountains' and identify $V \approx D^2h$).
}

{
The `emergent kinematic viscosity' $\nu$ of the superfluid's turbulent 
layer can therefore be estimated from the balance
between the energy (per unit mass per unit time),
$3 v^3/(8h)$ which is fed into the turbulence by the
incoming flow, and the energy (per unit mass per unit time) 
$\nu \kappa^2 {\cal L}^2$
which is dissipated into waves. 
It is convenient to express  
$\nu$ in terms of the quantum of circulation $\kappa$. 
%(which is equal to $2 \pi$ in our units based on $\xi$ and $c$).
We obtain,
}
\begin{equation}
\frac{\nu}{\kappa}= \frac{3 v^3}{8 \kappa^3 D {\cal L}^2}.
\end{equation}

\noindent
{
In our dimensionless units based on $\xi$ and $c$, the quantum of circulation is
$\kappa=2 \pi$. At the three applied flow speeds
we find that the saturated turbulent layer has {length} (see Fig.~3)
$L \approx 2000 \xi, 7000 \xi$ and $15000 \xi$ corresponding
to $\nu/\kappa =2.4, 1.5$ and $1.1$ respectively. 
}

{These values are larger than
$\nu/\kappa \approx 0.1$
reported \cite{Golov2008} in experimental studies of the
`Vinen' (or `ultraquantum') turbulent regime in He$^4$. 
We argue that the explanation of this difference is that
in the helium experiments the vortex lines are relatively
farther away from each other.
Since dissipation arises from sound emission from rapidly accelerating
vortices and from vortex reconnections, we expect 
more dissipation in our problem.}

{To quantify the
relative separation between the vortex lines, we compare
the average intervortex distance $\delta \approx {\cal L}^{-1/2}$
to the vortex core radius $a_0$,
which is $a_0 \approx 10^{-10}~\rm m$
in helium and $a_0 \approx 5 \xi$ in our problem.
Clearly, at the smallest distance between two vortex lines,  
$\delta \approx a_0$, the two
vortices either rotate around each other with speed
approaching the speed of sound, or undergo a vortex reconnection,
turning kinetic into acoustic energy at the highest rate.}

{
Figure 3 of Ref.~\cite{Golov2008}
shows vortex line density ${\cal L} \approx 2 \times 10^7~\rm m^{-2}$, 
which is typical of other He$^4$ experiments, yielding
$\delta/a_0 \approx 2 \times 10^6$. 
In our problem, corresponding to the three imposed speeds, we find
$\delta /a_0 \approx 13.8, 7.4$ and $5.1$: the vortex lines
are more closely packed, therefore it is not unexpected that in
our problem the 
dissipation is larger.
}


\begin{thebibliography}{99}

%%%%%%%%%%%%%%%%%%%%%%%%%%%%%%%%%%%%%%%%%%%%%%%%%%%%%%%%%%%
% introduction
\bibitem{Barenghi-PNAS}
C. F. Barenghi, L. Skrbek, and K. R. Sreenivasan,
%Introduction to quantum turbulence.
{\it Proc. Nat. Acad. Sci. USA}, {\bf 111} (Suppl. 1), 4647 (2014).

%%%%%%%%%%%%%%%%%%%%%%%%%%%%%%%%%%
% more intro

\bibitem{Bradley2011-NatPhys}
D. I. Bradley, S. N. Fisher, A. M. Gu\'{e}nault, R. P. Haley, G. R.
Pickett, G. Potts, and V. Tsepelin,
%Direct measurement of energy dissipated by quantum turbulence.
{\it Nature Phys.} {\bf 7}, 473 (2011).

\bibitem{Zmeev2015a}
D. E. Zmeev, P. M. Walmsley, A. I. Golov, P. V. E. McClintock, S. N.
Fisher and W. F. Vinen,
%Dissipation of quasiclassical turbulence in superfluid $^4$He.
{\it Phys. Rev. Lett.}, {\bf 115}, 155303 (2015).

\bibitem{Boue2013}
L. Bou\'{e}, V. L'vov, A. Pomyalov and I. Procaccia,
%Enhancement of intermittency in superfluid turbulence.
{\it Phys. Rev. Lett.} {\bf 110}, 014502 (2013).

%%%%%%%%%%%%%%%%%%%%%%%%%%%%%%%%%%%%%%%%%%%%%%%%%%%%%%%%%%%
% grids
\bibitem{Davis2000} 
S. I. Davis, P. C. Hendry and  P. V. E. McClintock,
%Decay of quantized vorticity in superfluid $^4$He at mK temperatures.
{\it Physica B (Cond. Matt.)} {\bf 280}, 43 (2000).

%%%%%%%%%%%%%%%%%%%%%%%%%%%%%%%%%%%%%%%%%%%%%%%%%%%%%%%%%%%
%wires
\bibitem{Guenault1986} 
A. M. Gu\'{e}nault, V. Keith, C. J.  Kennedy,
S. G. Mussett and  G. R. Pickett,
%The mechanical behavior of a vibrating wire in superfluid 3He-B
%in the ballistic limit.
%{\it J. Low Temp. Phys.} {\bf 62} 511-523 (1986).
{\it J. Low Temp. Phys.} {\bf 62}, 511 (1986).


\bibitem{Bradley2005}
D. I. Bradley, D. O. Clubb, S. N. Fisher,
A. M. Gu\'{e}nault, R. P. Haley,  C. J. Matthews, 
G. R. Pickett, V. Tsepelin  and K. Zaki,
%Emission of discrete vortex rings by a vibrating grid 
%in superfluid 3He-B: a precursor to quantum turbulence.
{\it Phys. Rev. Lett.} {\bf 95}, 035302 (2005).

\bibitem{Bradley2011}
D. I. Bradley, M. \v{C}love\v{c}ko, M. J. Fear,
S. N. Fisher, A. M. Gu\'{e}nault,
R. P. Haley, C. R. Lawson, G. R. Pickett,
R. Schanen, V. Tsepelin and P. Williams,
%A new device or studying low or zero frequency mechanical motion
%at very low temperatures.
%{\it J. Low Temp. Phys.}  {\bf 165}, 114-131 (2011).
{\it J. Low Temp. Phys.}  {\bf 165}, 114 (2011).

\bibitem{Fisher2001}
S. N. Fisher,  A. J. Hale, A. M. Gu\'{e}nault
and G. R. Pickett,
%Generation and Detection of Quantum Turbulence in Superfluid 3He-B.
{\it Phys. Rev. Lett.} {\bf 86}, 244 (2001).

%%%%%%%%%%%%%%%%%%%%%%%%%%%%%%%%%%%%%%%%%%%%%%%%%%%%%%%%%%%%%%%
%forks
\bibitem{Blaauwgeers2007}
R. Blaauwgeers,  M. Blazkova, M. \v{C}love\v{c}ko,
V. B. Eltsov  R. de Graaf, J. 
Hosio, M. Krusius, D. Schmoranzer, W. Schoepe,
L. Skrbek, P. Skyba,
R. E. Solntsev and D. E. Zmeev,
%Quartz tuning fork: thermometer, pressure and viscometer for helium liquids.
{\it J. Low Temp. Phys.} {\bf 146}, 537 (2007).

\bibitem{Bradley2012}
D. I. Bradley, M. \v{C}love\v{c}ko, S. N. Fisher, D.
Garg, E. Guise, R. P. Haley, O. 
Kolosov, G. R. Pickett, V. Tsepelin,
D. Schmoranzer and L. Skrbek,
%Crossover from hydrodynamic to acoustic drag on quartz tuning forks in normal and superfluid 4He.
{\it Phys. Rev. B} {\bf 85}, 014501 (2012).

%%%%%%%%%%%%%%%%%%%%%%%%%%%%%%%%%%%%%%%%%%%%%%%%%%%%%%%%%%%
% propeller
\bibitem{Tabeling1998}
J. Maurer and P. Tabeling, 
%Local investigation of superfluid turbulence.
%{\it Europhys. Lett.} {\bf 43}, 29-34 (1998)
{\it Europhys. Lett.} {\bf 43}, 29 (1998)


\bibitem{Salort2010}
J. Salort, C. Baudet, B. Castaing, B. Chabaud,
F. Davidaud, T. Didelot,
P. Diribarne, B. Dubrulle, Y. Cagne and F. Gauthier,
%Turbulent velocity spectra in superfluid flows.
{\it Phys. Fluids} {\bf 22}, 125102 (2010);

%%%%%%%%%%%%%%%%%%%%%%%%%%%%%%%%%%%%%%%%%%%%%%%%%%%%%%%%%%%
%spheres
\bibitem{Schoepe1995}
J. J\"{a}ger, B. Schudurer and W. Schoepe, 
%Turbulent and laminar drag of superfluid helium on an oscillating microsphere.
{\it Phys. Rev. Lett.} {\bf 74}, 566 (1995).

%%%%%%%%%%%%%%%%%%%%%%%%%%%%%%%%%%%%%%%%%%%%%%%%%%%%%%%%%%%%%%%%%
% moving objects
\bibitem{VinenSkrbek2008}
W. F. Vinen and L. Skrbek,
%The use of vibrating structures in the study of quantum turbulence. 
{\it Progress in Low Temperature Physics},
ed. by W.P. Halperin and M. Tsubota, Elsevier (Amsterdam),
Vol XVI, Chap 4, pp 195-246 (2008).

%%%%%%%%%%%%%%%%%%%%%%%%%%%%%%%%%%%%%%%%%%%%%%%%%%%%%%%%%%%%%%%%%%
% visualization

\bibitem{Zmeev2015b}
D. E. Zmeev, F. Pakpour,  P. M. Walmsley,
A. I. Golov, W. Guo, D. N. McKinsey,
G. G. Ihas, P. V. E. McClintock,
S. N. Fisher and W. F. Vinen,
%Excimers He$^*_2$ as tracers of quantum turbulence in He4 in the T=0 limit.
{\it Phys. Rev. Lett.} {\bf 110}, 175303 (2013).

\bibitem{Duda2015}
D. Duda, P. \u{S}van\u{c}ara, M. La Mantia, M. Rotter
and L. Skrbek,
%Visualization of viscous and quantum flows of liquid He4 due to an 
%oscillating cylinder of rectangular cross section.
{\it Phys. Rev. B} {\bf 92}, 064519 (2015).

\bibitem{Lathrop}
G. P. Bewley, D. P. Lathrop and K. R. Sreenivasan,  
%Visualization of quantized vortices
{\it Nature} {\bf 441}, 588 (2006).

\bibitem{Xu2007}
 {T. Xu and S. W. Van Sciver,
%Particle image velocimetry measurements of the velocity profile
%om He~II forced flow,
{\it Phys. Fluids} {\bf 19}, 071703 (2007).
}

\bibitem{Guo2014}
 {W. Guo, M. La Mantia, D. P. Lathrop and S. W. Van Sciver,
%Visualization of two-fluid flows of superfluid helium-4,
{\it Proc. Nat. Aca. Sci. USA} {\bf 111}, Suppl.~1, 4653 (2014).
}
%%%%%%%%%%%%%%%%%%%%%%%%%%%%%%%%%%%%%%%%%%%%%%%%%%%%%%%%%
% nucleation
\bibitem{Pomeau1992}
T. Frisch, Y. Pomeau, and S. Rica,
%Transition to dissipation in a model of superflow,
{\it Phys. Rev. Lett.} {\bf 69}, 1644 (1992).

%%%%%%%%%%%%%%%%%%%%%%%%%%%%%%%%%%%%%%%%%%%%%%%%%%%%%%%%%%%
% vortex mill

\bibitem{Schwarz-mill}
K. W. Schwarz,
%Phase slip and turbulence in superfluid $^4$He: a vortex mill that works.
{\it Phys. Rev. Lett.} {\bf 64}, 1130 (1990).

%%%%%%%%%%%%%%%%%%%%%%%%%%%%%%%%%%%%%%%%%%%%%%%%%%%%%%%%%%
% pinning
\bibitem{Schwarz-1981-pinning}
K. W. Schwarz,
%Vortex pinning in superfluid helium.
{\it Phys. Rev. Lett.} {\bf 47}, 251 (1981).

%\bibitem{Tsubota-1993-pinning}
%M. Tsubota and S. Maekawa,
%Pinning and depinning of two quantized vortices in superfluid $^4$He.
%Phys. Rev. B {\bf 47}, 12040 (1993).

\bibitem{Tsubota-1994-pinning}
M. Tsubota,
%Capacity of a pinning site for trapping quantized vortices
%in superfluid $^4$He.
{\it Phys. Rev. B} {\bf 50}, 579 (1994).


%%%%%%%%%%%%%%%%%%%%%%%%%%%%%%%%%%%%%%%%%%%%%%%%%%%%%%%%%%%
% GPE simulation of flow past a hemisphere on a plane
\bibitem{winiecki}
T. Winiecki, PhD thesis, University of Durham (2001).

%%%%%%%%%%%%%%%%%%%%%%%%%%%%%%%%%%%%%%%%%%%%%%%%%%%%%%%%%%%

% ellipse
\bibitem{Stagg-ellipse} 
G. W. Stagg, N. G. Parker and  C. F. Barenghi,  
%Quantum analogues of classical wakes in Bose-Einstein condensates.
{\it J. Phys. B} {\bf 47}, 095304 (2014).

%%%%%%%%%%%%%%%%%%%%%%%%%%%%%%%%%%%%%%%%%%%%%%%%%%%%%%%%%%%
% remanent
\bibitem{Yano-2007}
N. Hashimoto, R. Goto, H. Yano, K. Obara,
O. Ishikawa and T. Hata,
%Control of turbulence in boundary layers of superfluid $^4$He
%by filtering out remanent vortices.
{\it Phys. Rev. B} {\bf 76}, 020504(R) (2007).

%%%%%%%%%%%%%%%%%%%%%%%%%%%%%%%%%%%%%%%%%%%%%%%%%%%%%%%%%%%
% AFM

\bibitem{Lawson}
C. R. Lawson, 
%{\it A novel measurement device for use in multiphase helium-3 and
%4 at ultra--low temperatures}. 
PhD thesis, Lancaster University (2013).

%%%%%%%%%%%%%%%%%%%%%%%%%%%%%%%%%%%%%%%%%%%%%%%%%%%%%%%%%%%%%
% GPE
%\bibitem{RobertsBerloff-GPE}
%P. H. Roberts and  N. G. Berloff,
%The nonlinear Schroedinger equation as a model of superfluidity.
%in {\it Quantized vortex dynamics and superfluid turbulence},
%eds C. F. Barenghi, R. J. Donnelly and W. F. Vinen, Springer (Berlin, Heidelberg)
%(2001), page 235--257.

\bibitem{SM}  See Supplemental Material section at the end of this manuscript.

%%%%%%%%%%%%%%%%%%%%%%%%
%% Origin of GPE
%\bibitem{pitaevskii_gross}
%Ginzburg, V. L. and Pitaevskii, L. P. {\it Zh. Eksp. Teor. Fiz.} {\bf 34}, 1240 (1958); Gross, E. P. {\it J. Math. Phys.} {\bf 4}, 195 (1963).

%%%%%%%%%%%%%%%%%%%%%%%%%%%%%%%%%%%%%%%%%%%%%%%%%%%%%%%%%%%%%%
% VFM
\bibitem{Schwarz1988}
K. W. Schwarz, %Three-dimensional vortex dynamics in superfluid 4He: Homogeneous
%superfluid turbulence. 
{\it Phys. Rev. B} {\bf 38}, 2398 (1988).

%%%%%%%%%%%%%%%%%%%%%%%%%%%%%%%%%%%%%%%%%%%%%%%%%%%%%%%%%%%
% sphere
\bibitem{Hanninen-sphere}
R. H\"{a}nninen, M. Tsubota and W. F. Vinen, 
% Generation of turbulence by oscillating structures 
%in superfluid helium at very low temperature.
{\it Phys. Rev. B} {\bf 75}, 064502 (2007).

\bibitem{Kivotides-sphere}
D. Kivotides, C. F. Barenghi and Y. A. Sergeev,
%Interactions between particles and quantized vortices in superfluid helium.
{\it Phys. Rev. B} {\bf 77}, 014527 (2008).

%%%%%%%%%%%%%%%%%%%%%%%%%%%%%%%%%%%%%%%%%%%%%%%%%%%%%%%%%%%
% cylinder

\bibitem{Hanninen-PNAS}
R. H\"{a}nninen and A. W. Baggaley, 
%Vortex filament method as a tool for computational
%visualization of quantum turbulence
{\it Proc. Nat. Acad. Sci USA} {\bf 111} (Suppl. 1), 4667 (2014).

\bibitem{Goto2008}
R. Goto, S. Fujiyama, H. Yano, Y. Nago,
N. Hashimoto, K. Obara, O. Ishikawa,
M. Tsubota and T. Hata,
%Turbulence in boundary flow of
%superfluid 4He triggered by free vortex rings. 
{\it Phys. Rev. Lett.} {\bf 100}, 045301 (2008).





\bibitem{Kursa2011}
 {M. Kursa, K. Bajer and T. Lipniacki, 
%Cascade of vortex loops initiated
%by a single reconnection of quantum vortices,
{\it Phys. Rev. B} {\bf 83}, 014515 (2011).}

\bibitem{Kerr2011}
 {
R. M. Kerr, 
%Vortex stretching as a mechanism for quantum kinetic energy decay,
{\it Phys. Rev. Lett.} {\bf 106}, 224501 (2011).
}

\bibitem{Svistunov1995}
 {B. V. Svistunov, 
%Superfluid turbulence in the low temperature limit,
{\it Phys. Rev. B} {\bf  52}, 3647 (1995).}

\bibitem{vz}  
$\langle v \rangle $ is
the $x$-component of the velocity averaged over the $xy$ plane and
5 time steps in the saturated regime.



\bibitem{Parker2005}
{
C. F. Barenghi, N. G. Parker, N. P. Proukakis and C. S. Adams,
%Decay of quantised vorticity by sound emission,
{\it J. Low Temp. Phys.} {\bf 138}, 629 (2005).
}

\bibitem{Leadbeater2001}
{
M. Leadbeater, T. Winiecki, D.C. Samuels, C.F. Barenghi and C.S. Adams,
%Sound emission due to superfluid vortex reconnections,
{\it Phys. Rev. Lett.} {\bf 86}, 1410 (2001).
}

\bibitem{Golov2008}
{
P. M. Walmsley and A. I. Golov,
%Quantum and quasiclassical types of superfluid turbulence,
{\it Phys. Rev. Lett.} {\bf 100}, 245301 (2008).
}

\bibitem{Reeves2015}
{
M. T. Reeves, T. P. Billam, B. P. Anderson and A. S. Bradley,
%Identifying a superfluid Reynolds number via dynamical similarity
{\it Phys. Rev. Lett.} {\bf 114}, 155302 (2015.
}

\bibitem{Schoepe2015}
{W. Schoepe,
% Superfluid Reynolds number and the transition from potential flow
%to turbulence in superfluid He$^4$ at milliKelvin temperatures
{\it JETP Lett.} {\bf 102}, 105 (2015).
}

\bibitem{Sasaki2010}
{K. Sasaki, N. Suzuki, and H. Saito,
{\it Phys. Rev. Lett.} {\bf 104}, 150404 (2010).
}

\bibitem{Kwon2016}
{W. J. Kwon, J. H. Kim, S. W. Seo and Y. Shin,
{\it Phys. Rev. Lett.} {\bf 117}, 245301 (2016).
}


\bibitem{data}  DOI to be provided once accepted.

%%% REFS IN SUPP MAT
\bibitem{Pethick} ``Bose-Einstein Condensation in Dilute Gases", C. J. Pethick and H. Smith (Cambrige University Press, Cambridge, 2002)

\bibitem{RobertsBerloff-GPE}
{
P. H. Roberts and  N. G. Berloff,
%The nonlinear Schroedinger equation as a model of superfluidity.
in {\it Quantized vortex dynamics and superfluid turbulence},
eds C. F. Barenghi, R. J. Donnelly and W. F. Vinen, Springer 
(Berlin, Heidelberg) (2001), page 235--257.
}
\bibitem{Rayfield1964}
Rayfield, G.~W., and Reif, F.~D.
%Quantized Vortex Rings in Superfluid Helium
{\it Phys. Rev. A} {\bf 136}, A1194 (1964).

\bibitem{Primer} ``A Primer in Quantum Fluids", C. F. Barenghi and N. G. Parker (Springer, Berlin, 2016)

\end{thebibliography}
\end{document}